\documentclass{article}

\usepackage{amsmath, amssymb}

\newcommand{\ZM}{\mathbb{Z}}

\newcommand{\CM}{\mathbb{C}}
\newtheorem{theorem}{Theorem}
 
\newtheorem{prop}{Proposition} 
\newtheorem{cor}{Corollary}

\newcommand{\bec}[1]{\mbox{\boldmath $#1$}}

\begin{document}

\title{{\bf Absolute zeta functions \\ 
for zeta functions of quantum walks \\ }
\vspace{15mm}}

\author{Jir\^o AKAHORI, $\quad$ Norio KONNO, $\quad$ Rikuki OKAMOTO$^{\ast}$ \\ \\
Department of Mathematical Sciences \\
College of Science and Engineering \\
Ritsumeikan University \\
1-1-1 Noji-higashi, Kusatsu, 525-8577, JAPAN \\
e-mail: akahori@se.ritsumei.ac.jp, \  n-konno@fc.ritsumei.ac.jp$^{\ast}$ \\
\\
\\
Iwao SATO \\ \\
Oyama National College of Technology \\
771 Nakakuki, Oyama 323-0806, JAPAN \\
e-mail: isato@oyama-ct.ac.jp 
}

\date{\empty }

\maketitle

\vspace{50mm}


\vspace{20mm}


\begin{small}
\par\noindent
{\bf Corresponding author}: Rikuki Okamoto, Department of Mathematical Sciences, College of Science and Engineering, Ritsumeikan University, 1-1-1 Noji-higashi, Kusatsu, 525-8577, JAPAN \\
e-mail: ra0099vs@ed.ritsumei.ac.jp
\par\noindent
\\
\par\noindent
{\bf Abbr. title: Absolute zeta functions for quantum walks} 
\end{small}









\clearpage

\begin{abstract}
This paper presents a connection between the quantum walk and the absolute mathematics. The quantum walk is a quantum counterpart of the classical random walk. We especially deal with the Grover walk on a graph. The Grover walk is a typical model of quantum walks. The time evolution of the Grover walk is obtained by a unitary matrix that is called the Grover matrix. We define the zeta function determined by the Grover matrix. First we prove that the zeta function of the Grover walk is the absolute automorphic form that constructs the absolute zeta function. Next we calculate the absolute zeta function defined by Grover walks on some graphs. The absolute zeta functions of the Grover walks are expressed by the multiple gamma function. Other types of absolute zeta functions are obtained as an analogue of the multiple gamma function.
\end{abstract}

\vspace{10mm}

\begin{small}
\par\noindent
{\bf Keywords}: Quantum walk, Absolute zeta function, Grover walk, Konno-Sato theorem
\end{small}

\vspace{10mm}

\section{Introduction \label{sec01}}
\quad This work is a continuation of our previous papers \cite{AkahoriEtAL2023, Konno2023}. The quantum walk (QW) has multiple states called chirality. Then the QW causes dynamics by a unitary matrix. In that sense, QW is a non-commutative modeling of the classical random walk (RW). Concerning QW, see \cite{GZ, Konno2008, ManouchehriWang, Portugal, Venegas}, and as for RW, see \cite{Norris, Spitzer}, for instance. This paper deals with the Grover walk on the simple connected graph $G$. The Grover walk is a well-studied model in QW research. Let $\mathbf{U}_G$ be the time evolution matrix of the Grover walk. In this paper, $\mathbf{U}_G$ is called the Grover matrix.  Then we define a zeta function $\zeta_{\mathbf{U}_G}$ of the Grover walk on $G$.\\
\quad On the other hand, the absolute zeta function is considered as the zeta function over the absolute field $\mathbb{F}_1$. Here $\mathbb{F}_1$ is the algebra of the absolute mathematics constructed and developed by mainly Kurokawa and his coworkers.  Concerning the absolute mathematics, see \cite{CC, KF1, Kurokawa3, Kurokawa, KO, KT3, KT4, Soule}. The absolute zeta function is given by the function that satisfies the absolute automorphic property. Then we see that the zeta function $\zeta_{\mathbf{U}_G}$ of the Grover walk on $G$ is the absolute automorphic form. Therefore we can get the absolute zeta function $\zeta_{\zeta_{\mathbf{U}_G}}$ determined by the Grover matrix $\mathbf{U}_G$.\\ 
\quad The rest of this paper is organized as follows. In Section 2, we treat the definition of the QW and the Konno-Sato theorem. Section 3 briefly describes the absolute mathematics. We especially introduce the absolute zeta function. Section 4 explains a connection between the QW and the absolute zeta function. In Section 5, we give examples for some graphs. We calculate the absolute zeta function of QW by two ways. One is computed by the cyclotomic polynomials and then the absolute zeta is represented by the multiple gamma function. In the other case, we 
 directly compute the absolute zeta function by its definition and series expansion. Finally, Section 6 is devoted to conclusion.

\section{QW on  graph and Konno-Sato theorem \label{sec03}}
This section deals with the definition of QW and a relation between QW and zeta function. Moreover we consider the Konno-Sato theorem presented by \cite{KonnoSato}. This theorem treats a connection for eigenvalues between QWs and RWs. More specifically, the Grover walk (which is a QW determined by the Grover matrix) with flip-flop shift type (called F-type) and simple symmetric RW (whose walker jumps to each of its nearest neighbors with equal probability) on a graph. We assume that all graphs are simple and connected.

Let $G=(V(G),E(G))$ be a connected graph without multiple edges and loops. We define $V(G)$ as the set of vertices and $E(G)$ as the set of unoriented edges $uv$ joining two vertices $u$ and $v$. Moreover, let $n=|V(G)|$ and $m=|E(G)|$ be the number of vertices and edges of $G$, respectively. For $uv \in E(G)$, an arc $(u,v)$ is the oriented edge from $u$ to $v$. Let $D(G)$ be the set of oriented edges $(u,v)$, i.e., $D(G)= \{ (u,v),(v,u) \mid uv \in E(G) \}$. Then $D_G$ denotes the symmetric digraph of $G$. For $e=(u,v) \in D(G)$, set $u=o(e)$ and $v=t(e)$. Furthermore, let $e^{-1}=(v,u)$ be the {\em inverse} of $e=(u,v)$. For $v \in V(G)$, the {\em degree} $\deg {}_G \ v = \deg v = d_v $ of $v$ is the number of vertices adjacent to $v$ in $G$. If $ \deg {}_G \ v=k$ (constant) for each $v \in V(G)$, then $G$ is called {\em $k$-regular}. A {\em path $P$ of length $n$} in $G$ is a sequence $P=(e_1, \ldots ,e_n )$ of $n$ arcs such that $e_i \in D(G)$, $t( e_i )=o( e_{i+1} ) \ (1 \leq i \leq n-1)$. If $e_i =( v_{i-1} , v_i )$ for $i=1 , \cdots , n$ , then we write $P=(v_0, v_1, \ldots ,v_{n-1}, v_n )$. Put $ \mid P \mid =n$, $o(P)=o( e_1 )$ and $t(P)=t( e_n )$. Also, $P$ is called an {\em $(o(P),t(P))$-path}. We say that a path $P=( e_1 , \ldots , e_n )$ has a {\em backtracking} if $ e^{-1}_{i+1} =e_i $ for some $i \ (1 \leq i \leq n-1)$. A $(v, w)$-path is called a {\em $v$-cycle} (or {\em $v$-closed path}) if $v=w$. Let $B^r$ be the cycle obtained by going $r$ times around a cycle $B$. Such a cycle is called a {\em multiple} of $B$. A cycle $C$ is {\em reduced} if both $C$ and $C^2 $ have no backtracking.

Here we introduce the  {\em Ihara zeta} function which is an important function in the graph theory. The {\em Ihara zeta} function of a graph $G$ is a function of a complex variable $u$ with $|u|$ sufficiently small, defined by 
\begin{align*}
{\bf Z} (G, u)= \exp \left( \sum^{\infty}_{r=1} \frac{N_r}{r} u^r \right), 
\end{align*}
where $N_r$ is the number of reduced cycles of length $r$ in $G$. The {\em Ihara zeta} function has various equivalent representations. One of the expressions called {\em Ihara expression} is due to the adjacency matrix and the diagonal matrix of the degree of $G$. Let $G$ be a simple connected graph with $n$ vertices $v_1, \ldots ,v_n $. The {\em adjacency matrix} ${\bf A}_n = [a_{ij} ]$ is the $n \times n$ matrix such that $a_{ij} =1$ if $v_i$ and $v_j$ are adjacent, and $a_{ij} =0$ otherwise. The following result was obtained by Ihara \cite{Ihara} and Bass \cite{Bass}. 
\begin{theorem}
Let $G$ be a simple connected graph with $V(G)= \{ v_1 , \ldots , v_n \}$ and $m$ edges. Then we have
\begin{align*}
{\bf Z} (G,u )^{-1} =(1- u^2 )^{\gamma-1} 
\det \left( {\bf I}_n -u {\bf A}_n + u^2 ( {\bf D}_n - {\bf I}_n ) \right). 
\end{align*}
Here $\gamma$ is the Betti number of $G$ {\rm (}i.e., $\gamma = m - n +1${\rm )}, ${\bf I}_n$ is the $n \times n$ identity matrix, and ${\bf D}_n = [d_{ij}]$ is the $n \times n$ diagonal matrix with $d_{ii} = \deg v_i$ and $d_{ij} =0 \ (i \neq j)$. This expression is called Ihara expression.
\end{theorem}

The {\em Ihara zeta} function has a representation related to Grover walks on the graph \cite{RenEtAl}. Let $G$ be a simple connected graph with $V(G)= \{ v_1 , \ldots , v_n \}$ and $m$ edges. Set $d_j = d_{v_j} = \deg v_j \ (j=1, \ldots , n)$. Then the $2m \times 2m$ {\em Grover matrix} ${\bf U}_{2m} = [ U_{ef} ]_{e,f \in D(G)} $ of $G$ is defined by 
\begin{align}
U_{ef} =\left\{
\begin{array}{ll}
2/d_{t(f)} (=2/d_{o(e)} ) & \mbox{if $t(f)=o(e)$ and $f \neq e^{-1} $, } \\
2/d_{t(f)} -1 & \mbox{if $f= e^{-1} $, } \\
0 & \mbox{otherwise. }
\end{array}
\right. 
\label{real}
\end{align}
The discrete-time QW with the Grover matrix ${\bf U}_{2m}$ as a time evolution matrix is the Grover walk with F-type on $G$. Here we introduce the {\em positive support} ${\bf F}^+ = [ F^+_{ij} ]$ of a real matrix ${\bf F} = [ F_{ij} ]$ as follows: 
\begin{align*}
F^+_{ij} =\left\{
\begin{array}{ll}
1 & \mbox{if $F_{ij} >0$, } \\
0 & \mbox{otherwise}.
\end{array}
\right.
\end{align*}
Ren et al. \cite{RenEtAl} showed that the Perron-Frobenius operator (or edge matrix) of a graph is the positive support $({}^{\rm{T}}{\bf U}_{2m})^+ $ of the transpose of its Grover matrix ${\bf U}_{2m}$, i.e., 
\begin{align}
{\bf Z} (G,u)^{-1} = \det \left( {\bf I}_{2m} -u( {}^{\rm{T}}{\bf U}_{2m})^+ \right)= \det \left( {\bf I}_{2m} -u {\bf U}_{2m} ^+ \right). 
\label{toruko01}
\end{align}
The Ihara zeta function of a graph $G$ is just a zeta function on the positive support of the Grover matrix of $G$. That is, the {\em Ihara zeta} function corresponds to the positive support version of the Grover walk (defined by the positive support ${\bf U}_{2m} ^+$ of the Grover matrix with F-type) on $G$.

Now we propose another zeta function of a graph. Let $G$ be a simple connected graph with $m$ edges. Then we define a zeta function $ \overline{{\bf Z}} (G, u)$ of $G$ satisfying 
\begin{align}
\overline{{\bf Z}} (G, u)^{-1} = \det \left( {\bf I}_{2m} -u {\bf U}_{2m} \right).    
\label{toruko02}
\end{align}
In other words, this zeta function corresponds to the {\em Grover walk} (defined by the Grover matrix ${\bf U}_{2m}$) with F-type on $G$.

Moreover the $n \times n$ matrix ${\bf P}_{n} = [ P_{uv} ]_{u,v \in V(G)}$ is given by
\begin{align*}
P_{uv} =\left\{
\begin{array}{ll}
1/( \deg {}_G \ u)  & \mbox{if $(u,v) \in D(G)$, } \\
0 & \mbox{otherwise.}
\end{array}
\right.
\end{align*}
Note that the matrix ${\bf P}_n$ is the transition probability matrix of the simple symmetric RW on $G$.
In this setting, Konno and Sato \cite{KonnoSato} presented the following result which is called the {\em Konno-Sato} theorem. 

\begin{theorem}
\label{K-S}
Let $G$ be a simple connected graph with $n$ vertices and $m$ edges. Then  
\begin{align}  
\overline{{\bf Z}} (G, u)^{-1} 
= \det ( {\bf I}_{2m} - u {\bf U}_{2m} )
=(1-u^2)^{m-n} \det \left( (1+u^2) {\bf I}_{n} -2u {\bf P}_n \right).
\label{wakatakakage1a}
\end{align}
\label{KS} 
\end{theorem}       
If we take $u = 1/\lambda$, then Eq. \eqref{wakatakakage1a} implies
\begin{align}
\det \left( \lambda {\bf I}_{2m} - {\bf U}_{2m} \right)
= ( \lambda {}^2 -1)^{m-n} \det \left( ( \lambda {}^2 +1) {\bf I}_n -2 \lambda {\bf P}_n \right).
\label{wakatakakage00}
\end{align}
Furthermore, Eq. \eqref{wakatakakage00} can be rewritten as 
\begin{align}
\det \left( \lambda {\bf I}_{2m} - {\bf U}_{2m} \right)= ( \lambda {}^2 -1)^{m-n} \times \prod_{ \lambda {}_{{\bf P}_n}  \in {\rm Spec} ({\bf P}_n)} \left( \lambda {}^2 +1 -2 \lambda {}_{{\bf P}_n} \lambda \right), 
\label{star01}
\end{align}
where ${\rm Spec} ({\bf B})$ is the set of eigenvalues of a square matrix ${\bf B}$. More precisely, we also use the following notation: 
\begin{align*}
{\rm Spec} ({\bf B}) = \left\{ \left[ \lambda_1 \right]^{l_1}, \ \left[ \lambda_2 \right]^{l_2}, \ \ldots \ , \left[ \lambda_k \right]^{l_k} \right\},
\end{align*}
where $\lambda_j$ is the eigenvalue of ${\bf B}$ and $l_j \in \ZM_{>}$ is the multiplicity of $\lambda_j$ for $j=1,2, \ldots, k$. Set $|{\rm Spec} ({\bf B}) | = l_1 + l_2 + \cdots + l_k$. It follows from $\lambda {}^2 +1 -2 \lambda {}_{{\bf P}_n} \lambda =0$ that $\lambda{}_{{\bf U}_{2m}} \in {\rm Spec} ({\bf U}_{2m})$ is given by
\begin{align}
\lambda{}_{{\bf U}_{2m}} = \lambda {}_{{\bf P}_n} \pm i \sqrt{1- \lambda {}^2_{{\bf P}_n}}.
\label{star02}
\end{align}
Remark that $\lambda {}_{{\bf P}_n} \in [-1,1]$. Noting Eqs. \eqref{star01} and \eqref{star02}, we introduce ${\rm Spec} \left( {\bf U}_{2m} : {\rm RW} \right)$ and ${\rm Spec} \left( {\bf U}_{2m} : {\rm RW}^c \right)$ as follows:  
\begin{align*}
&{\rm Spec} \left( {\bf U}_{2m} : {\rm RW} \right) 
\\
& \qquad = \left\{ \left[ \lambda {}_{{\bf P}_n} + i \sqrt{1- \lambda {}^2_{{\bf P}_n}} \right]^1,  \ \left[ \lambda {}_{{\bf P}_n} - i \sqrt{1- \lambda {}^2_{{\bf P}_n}} \right]^1  \ : \  \lambda {}_{{\bf P}_n} \in {\rm Spec} ({\bf P}_n) \right\},
\\
&{\rm Spec} \left( {\bf U}_{2m} : {\rm RW}^c \right) 
= \left\{ \left[ 1 \right]^{|m-n|},  \ \left[ -1 \right]^{|m-n|} \right\}.
\end{align*}
When $m=n$, we let ${\rm Spec} \left( {\bf U}_{2m} : {\rm RW}^c \right) = \emptyset$. Note that $|{\rm Spec} \left( {\bf U}_{2m} : {\rm RW} \right)| = 2n$ and $|{\rm Spec} \left( {\bf U}_{2m} : {\rm RW}^c \right)| = 2 |m-n|$. We should remark that ${\rm Spec} \left( {\bf U}_{2m} : {\rm RW} \right)$ corresponds to the eigenvalue of ${\bf P}_n$ which is the transition probability matrix of the simple symmetric RW on $G$. On the other hand, ${\rm Spec} \left( {\bf U}_{2m} : {\rm RW}^c \right)$ does not corresponds to the RW, so superscript ``c" (of ${\rm RW}^c$) stands for ``complement". Therefore we obtain

\begin{cor}
Let $G$ be a simple connected graph with $n$ vertices and $m$ edges. 
\par
\
\par\noindent
{\rm (i)} If $m > n$, then   
\begin{align*}
{\rm Spec} ({\bf U}_{2m}) = {\rm Spec} \left( {\bf U}_{2m} : {\rm RW} \right) \cup {\rm Spec} \left( {\bf U}_{2m} : {\rm RW}^c \right).
\end{align*}
\par\noindent
{\rm (ii)} If $m = n$, then   
\begin{align*}
{\rm Spec} ({\bf U}_{2m}) = {\rm Spec} \left( {\bf U}_{2m} : {\rm RW} \right).
\end{align*}
\par\noindent
{\rm (iii)} If $m < n$, then   
\begin{align*}
{\rm Spec} ({\bf U}_{2m}) = {\rm Spec} \left( {\bf U}_{2m} : {\rm RW} \right) \setminus {\rm Spec} \left( {\bf U}_{2m} : {\rm RW}^c \right).
\end{align*}
\label{moderuna011}
\end{cor}
\par
\
\par
In Section \ref{sec04}, we will give some examples for each case. Since ${\bf U}_{2m}$ is unitary, each eigenvalue $\lambda \in {\rm Spec} ({\bf U}_{2m})$ satisfies $|\lambda|=1$, and so we put $\lambda = e^{i \theta} \ (\theta \in [0, 2 \pi))$. Thus we get
\begin{align}
\cos \theta = \frac{\lambda + \overline{\lambda}}{2},
\label{wakatakakage01}
\end{align}
where $\overline{\lambda}$ is also the complex conjugate of $\lambda \in \CM.$ This is called the Joukowsky transform. It follows from  Eq. \eqref{wakatakakage01} and $\lambda^{-1} = \overline{\lambda}$ that Eq. \eqref{wakatakakage00} becomes 
\begin{align*}
\det \left( \lambda {\bf I}_{2m} - {\bf U}_{2m} \right)
= ( \lambda {}^2 -1 )^{m-n} ( 2 \lambda)^{n} \det \left( \cos \theta \cdot {\bf I}_n - {\bf P}_n \right).
\end{align*}
Therefore we have a relation ``$\cos \theta \in {\rm Spec} ({\bf P}_n)$'' \ $\Longrightarrow$ \ ``$\lambda = e^{i \theta} \in {\rm Spec} ({\bf U}_{2m})$'' which is sometimes called the {\em spectral mapping theorem} in the study of QW (see \cite{SS}, for example). We will explain in a more detailed fashion. We set
\begin{align*}
{\rm Spec} ({\bf P}_n) = \left\{ \left[ \cos \theta_1 \right]^{l_1}, \ \left[ \cos \theta_2 \right]^{l_2}, \ \ldots \ , \left[ \cos \theta_p \right]^{l_p} \right\},
\end{align*}
where $|{\rm Spec} ({\bf P}_n) | = l_1 + l_2 + \cdots + l_p = n$ and $0 = \theta_1 < \theta_2 < \cdots < \theta_p < 2 \pi$. Then we get
\begin{align*}
&{\rm Spec} \left( {\bf U}_{2m} : {\rm RW} \right) 
\nonumber
\\
& \qquad = \left\{ \left[ e^{i \theta_1} \right]^{l_1}, \ \left[ e^{- i \theta_1} \right]^{l_1}, \ \left[ e^{i \theta_2} \right]^{l_2}, \ \left[ e^{- i \theta_2} \right]^{l_2}, \ \ldots \ , \left[ e^{i \theta_p} \right]^{l_p}, \ \left[ e^{- i \theta_p} \right]^{l_p} \right\}.
\end{align*}
Remark that $|{\rm Spec} \left( {\bf U}_{2m} : {\rm RW} \right) | = 2 \left( l_1 + l_2 + \cdots + l_p \right) = 2n$. 

\section{Absolute zeta function \label{sec02}}
First we introduce the following notation: $\mathbb{Z}$ is the set of integers, $\mathbb{Z}_{>} = \{1,2,3, \ldots \}$,  $\mathbb{R}$ is the set of real numbers, and $\mathbb{C}$ is the set of complex numbers. 

In this section, we briefly review the framework on the absolute automorphic forms, the absolute 
 Hurwitz zeta functions and the absolute zeta function which can be considered as the zeta function over $\mathbb{F}_1$ (see \cite{Kurokawa3, Kurokawa, KO, KT3, KT4} and references therein, for example). 

Let $f$ be a function $f : \mathbb{R} \to \mathbb{C} \cup \{ \infty \}$. We say that $f$ is an {\em absolute automorphic form} of weight $D$ if $f$ satisfies
\begin{align*}
f \left( \frac{1}{x} \right) = C x^{-D} f(x)
\end{align*}
with $C \in \{ -1, 1 \}$ and $D \in \mathbb{Z}$. The {\em absolute Hurwitz zeta function} $Z_f (w,s)$ is defined by
\begin{align*}
Z_f (w,s) = \frac{1}{\Gamma (w)} \int_{1}^{\infty} f(x) \ x^{-s-1} \left( \log x \right)^{w-1} dx,
\end{align*}
where $\Gamma (x)$ is the gamma function (see \cite{Andrews1999}, for instance). Then taking $x=e^t$, we see that $Z_f (w,s)$ can be rewritten as the Mellin transform: 
\begin{align}
Z_f (w,s) = \frac{1}{\Gamma (w)} \int_{0}^{\infty} f(e^t) \ e^{-st} \ t^{w-1} dt.
\label{kirishima01}
\end{align}
Moreover, we introduce the {\em absolute zeta function} $\zeta_f (s)$ as follows:
\begin{align*}
\zeta_f (s) = \exp \left( \frac{\partial}{\partial w} Z_f (w,s) \Big|_{w=0} \right).
\end{align*}
\par
Now we give an example of $f(x)$ which will be used in Section \ref{sec05}:
\begin{align}
f(x) = \frac{1}{(x^N -1)^2}
\label{kirishima02}
\end{align}
with $N \in \mathbb{Z}_{>}$. Then we find 
\begin{align*}
f \left( \frac{1}{x} \right) = x^{2N} f(x),
\end{align*}
so $f$ (given by Eq. \eqref{kirishima02}) is an absolute automorphic form of weight $-2N$. From Eq. \eqref{kirishima01}, we compute
\begin{align*}
Z_f (w,s) 
&= \frac{1}{\Gamma (w)} \int_{0}^{\infty} \frac{e^{-st}}{(e^{Nt} -1)^2} \ t^{w-1} dt
\\
&= \frac{1}{\Gamma (w)} \int_{0}^{\infty} e^{-st} \ \left( \sum_{n_1=1}^{\infty} e^{-n_1 N t} \right) \left( \sum_{n_2=1}^{\infty} e^{-n_2 N t} \right) \ t^{w-1}dt
\\
&= \sum_{n_1=0}^{\infty} \sum_{n_2=0}^{\infty} \int_{0}^{\infty} \frac{1}{\Gamma (w)} e^{- \{s + 2N + (n_1+n_2)N \}t} \ t^{w-1} dt
\\
&= \sum_{n_1=0}^{\infty} \sum_{n_2=0}^{\infty} \{s + 2N + (n_1+n_2)N \}^{-w}.
\end{align*}
Thus we have 
\begin{align}
Z_f (w,s) = \sum_{n_1=0}^{\infty} \sum_{n_2=0}^{\infty} \{s + 2N + (n_1+n_2)N \}^{-w}.
\label{kaibutsu01}
\end{align}
Here we introduce the {\em multiple Hurwitz zeta function of order $r$}, $\zeta_r (s, x, (\omega_1, \ldots, \omega_r))$, and the {\em multiple gamma function of order $r$}, $\Gamma_r (x, (\omega_1, \ldots, \omega_r))$, in the following (see \cite{Kurokawa3, Kurokawa, KT3}): 
\begin{align}
\zeta_r (s, x, (\omega_1, \ldots, \omega_r))
&= \sum_{n_1=0}^{\infty} \cdots \sum_{n_r=0}^{\infty} \left( n_1 \omega_1 + \cdots + n_r \omega_r + x \right)^{-s}, 
\label{kirishima03}
\\
\Gamma_r (x, (\omega_1, \ldots, \omega_r)) 
&= \exp \left( \frac{\partial}{\partial s} \zeta_r (s, x, (\omega_1, \ldots, \omega_r)) \Big|_{s=0} \right).
\label{kirishima04}
\end{align}
Therefore, combining Eq. \eqref{kaibutsu01} with Eqs. \eqref{kirishima03} and \eqref{kirishima04} yields
\begin{align}
Z_f (w,s) &= \zeta_2 (w, s+2N, (N,N)), 
\label{kirishima05}
\\
\zeta_f (s) &= \Gamma_2 (s+2N, (N,N)). 
\label{kirishima06}
\end{align}
So we see that $Z_f (w,s)$ and $\zeta_f (s)$ can be obtained by a multiple Hurwitz zeta function of order $2$ and a multiple gamma function of order $2$, respectively.

\section{Relation between QW and absolute zeta function \label{sec04}}
Inspired by Eqs. \eqref{toruko01} and \eqref{toruko02}, we consider a zeta function $\zeta_{{\bf A}} (u)$ for an $N \times N$ matrix ${\bf A}$ defined by
\begin{align*}  
\zeta_{{\bf A}} (u) = \left\{ \det \left( {\bf I}_{N} - u {\bf A} \right) \right\}^{-1}.
\end{align*}
If ${\bf A}$ is a regular matrix with its eigenvalues $\{\lambda_1, \ldots, \lambda_N\}$, then we easily see
\begin{align*}  
\zeta_{{\bf A}} \left( \frac{1}{u} \right)^{-1} 
&= \det \left( {\bf I}_{N} - \frac{1}{u} {\bf A} \right)
= \prod_{k=1}^N \left( 1 - \frac{\lambda_k}{u} \right)
= \left(\frac{1}{u} \right)^N \prod_{k=1}^N \left( u - \lambda_k \right)
\\
&
= \left(\frac{1}{u} \right)^N \left( \prod_{k=1}^N \lambda_k \right) \ (-1)^N \ \prod_{k=1}^N \left( 1 - \frac{u}{\lambda_k} \right)
\\
&
= (-u)^{-N} \ \det {\bf A} \ \left\{ \zeta_{{\bf A}^{-1}} \left( u \right) \right\}^{-1}. 
\end{align*}
Therefore we have the following result.
\begin{align}
\zeta_{{\bf A}} \left( \frac{1}{u} \right) = (-u)^{N} \ \left(\det {\bf A} \right)^{-1} \ \zeta_{{\bf A}^{-1}} \left( u \right).
\label{kakumei01}
\end{align}
On the other hand, we defined our zeta function $\overline{{\bf Z}} (G, u)$ based on a QW on the graph $G$ by
\begin{align}  
\overline{{\bf Z}} (G, u) = \left\{ \det \left( {\bf I}_{2m} - u {\bf U}_{2m} \right) \right\}^{-1},
\label{kakumei02}
\end{align}
where $G$ is a simple connected graph with $n$ vertices and $m$ edges. In order to clarify the dependence on a graph, from now on, we define ``${\bf U}_{2m}$ and $\overline{{\bf Z}} (G, u)$" by ``${\bf U}_{G}$ and $\zeta_{{\bf U}_{G}}$", respectively. So Eq. \eqref{kakumei02} can be rewritten as 
\begin{align*}  
\zeta_{{\bf U}_{G}} (u) = \left\{ \det \left( {\bf I}_{2m} - u {\bf U}_{G} \right) \right\}^{-1}.
\end{align*}
Then it follows from Eq. \eqref{kakumei01} that 
\begin{align}
\zeta_{{\bf U}_{G}} \left( \frac{1}{u} \right) = (-u)^{2m} \ \left(\det {\bf U}_{G} \right)^{-1} \ \zeta_{{\bf U}_{G}^{-1}} \left( u \right).
\label{kakumei04}
\end{align}
We see that the definition of ${\bf U}_{G} (={\bf U}_{2m})$ given by Eq. \eqref{real} implies that each component of ${\bf U}_{G}$ is a real number. So ${\bf U}_{G}$ becomes an orthogonal matrix. Thus we have
\begin{align} 
{\bf U}_{G} ^{-1} = {}^{\rm{T}}{\bf U}_{G},
\label{kakumei05}
\end{align}
where $\rm{T}$ is the transposed operator. From Eq. \eqref{kakumei05}, we compute
\begin{align*}  
\zeta_{{\bf U}_{G}} (u) 
&= \left\{ \det \left( {\bf I}_{2m} - u {\bf U}_{G} \right) \right\}^{-1} = \left\{ \det \left( {\bf I}_{2m} - u {}^{\rm{T}}{\bf U}_{G} \right) \right\}^{-1} 
\\
&= \left\{ \det \left( {\bf I}_{2m} - u {\bf U}_{G}^{-1} \right) \right\}^{-1} = \zeta_{{\bf U}_{G} ^{-1}} (u). 
\end{align*}
So we have
\begin{align}  
\zeta_{{\bf U}_{G}} (u) = \zeta_{{\bf U}_{G} ^{-1}} (u).
\label{kakumei06}
\end{align}
Moreover we get
\begin{align} 
\det {\bf U}_{G} \in \{-1,1\},
\label{kakumei07}
\end{align}
since ${\bf U}_{G}$ is an orthogonal matrix. Combining Eq. \eqref{kakumei04} with Eqs. \eqref{kakumei06} and \eqref{kakumei07}, we obtain the following result for our zeta function $\zeta_{{\bf U}_{G}} (u)$:
\begin{theorem}
We assume that $G$ is a simple connected graph with $n$ vertices and $m$ edges. Let ${\bf U}_{G} (={\bf U}_{2m})$ be a time evolution matrix of the Grover walk on $G$. Then we have
\begin{align*}
\zeta_{{\bf U}_{G}} \left( \frac{1}{u} \right) = \det {\bf U}_{G} \ u^{2m} \ \zeta_{{\bf U}_{G}} \left( u \right),
\end{align*}
where $\det {\bf U}_{G} \in \{-1,1\}$. 
\label{kakumei09}
\end{theorem}
Recall that $f$ is an absolute automorphic form of weight $D$ if $f$ satisfies
\begin{align*}
f \left( \frac{1}{u} \right) = C \ u^{-D} \ f(u)
\end{align*}
with $C \in \{ -1, 1 \}$ and $D \in \mathbb{Z}$. Therefore, from Theorem \ref{kakumei09}, we have an important property of our zeta function $\zeta_{{\bf U}_{G}} (u)$, that is, ``$\zeta_{{\bf U}_{G}} (u)$ is an absolute automorphic form of weight $- 2m$".

\section{Example \label{sec05}}

Let $G$ be a simple connected graph with $n$ vertices and $m$ edges. This section treats four classes of graphs $G$, i.e., (i) cycle graph case, (ii) star graph case, (iii) complete bipartite graph case, and (iv) complete graph case. Star graph is an important case of the class of complete bipartite graph. Therefore we especially deal with the star graph case. We donate ${\bf U}_{2m}$ by ${\bf U}_{G}$. Similarly, we put ${\bf P}_{G} = {\bf P}_{n}$. Note that if we apply the Konno-Sato Theorem (Theorem \ref{KS}) to ``${\rm Spec} \left( {\bf P}_{G} \right)$", then we obtain ``${\rm Spec} \left( {\bf U}_{G} \right)$".

From now on, we consider two methods in order to compute the absolute zeta function $\zeta_{{\bf U}_{G}} (u)$ for a graph $G$. One is based on the Konno-Sato theorem. First, we obtain the eigenvalue of the Grover matrix $\mathbf{U}_G$. Then we compute the absolute zeta function by its definition. The other is given by the following result based in Kurokawa \cite{Kurokawa}.
\begin{theorem}
\label{zettaisugaku01chap2}
For $\ell \in \mathbb{Z}, \ m(i) \in \mathbb{Z}_{>} \ (i=1, \ldots, a), \ n(j) \in \mathbb{Z}_{>} \ (i=1, \ldots, b)$, let 
\begin{align*}
f(x) = x^{\ell/2} \ \frac{\left( x^{m(1)} - 1 \right) \cdots \left( x^{m(a)} - 1 \right)}{\left( x^{n(1)} - 1 \right) \cdots \left( x^{n(b)} - 1 \right)}.
\end{align*}
Moreover we put 
\begin{align*}
\bec{n} 
&= \left( n(1), \ldots, n(b) \right), \quad |\bec{n}| = \sum_{j=1}^b n(j), \quad m \left( I \right) = \sum_{i \in I} m(i), \quad |I| = \sum_{i \in I} 1,
\\
{\rm deg} (f) 
&= \frac{\ell}{2} + \sum_{i=1}^a m(i) - \sum_{j=1}^b n(j) = \frac{\ell}{2} + \sum_{i=1}^a m(i) - |\bec{n}|.
\end{align*}
Then we have 
\begin{align}
Z_f (w, s) 
&= \sum_{I \subset \{1, \ldots, a \}} (-1)^{a - |I|} \ \zeta_b \left( w, s - \frac{\ell}{2} + |\bec{n}| - m \left( I \right), \bec{n} \right) \notag
\\
&= \sum_{I \subset \{1, \ldots, a \}} (-1)^{|I|} \ \zeta_b \left( w, s - {\rm deg} (f) + m \left( I \right), \bec{n} \right),
\label{mkusatsu01b}
\\
\zeta_f (s) 
&= \prod_{I \subset \{1, \ldots, a \}} \Gamma_b \left( s - \frac{\ell}{2} + |\bec{n}| - m \left( I \right), \bec{n} \right)^{ (-1)^{a - |I|}} \notag
\\
&= \prod_{I \subset \{1, \ldots, a \}} \Gamma_b \left( s  - {\rm deg} (f) + m \left( I \right), \bec{n} \right)^{ (-1)^{|I|}}.
\label{mkusatsu02b}
\end{align}
\end{theorem}

The key point of this theorem is that the function $f$ must be a rational function of the above form. We first calculate the zeta function $\zeta_{\mathbf{U}_G}$ of the Grover walk. If the $\zeta_{\mathbf{U}_G}$ is represented by such a rational function, then we choose a based on Theorem 4. If the $\zeta_{\mathbf{U}_G}$ is not such a rational function, then we compute the eigenvalue by the Konno-Sato theorem. The absolute zeta function is obtained with the eigenvalue of $\mathbf{U}_G$.

\par
\
\par
(i) Cycle graph case

Let $G=C_n$ be the cycle graph with $n$ vertices and $m=n$ edges.
Put $\xi_k = 2 \pi k/n$ for $k=1, \ldots , n$. Then we get
\begin{align*}
{\rm Spec} \left( {\bf P}_{C_n} \right) 
&={\rm Spec} \left( {\bf P}_n \right) 
= \left\{ [\cos \xi_k ]^{1} \ : \ k=1, \ldots , n \right\}, 
\\
{\rm Spec} \left( {\bf U}_{C_n} \right) 
&= {\rm Spec} \left( {\bf U}_{2n} \right)
= \left\{ [e^{i \xi_k} ]^{1}, \  [e^{- i \xi_k} ]^{1} \ : \ k=1, \ldots , n \right\}. 
\end{align*}

By Theorem \ref{KS}, we obtain
\begin{align*}
\zeta_{{\bf U}_{C_n}} \left( u \right) ^{-1} 
&= \overline{{\bf Z}} (C_n, u)^{-1} = \det ( {\bf I}_{2n} - u {\bf U}_{C_n} )
\\
&=(1-u^2)^{n-n} \det \left\{ (1+u^2) {\bf I}_{n} -2u {\bf P}_{C_n} \right\}
\\
&= \prod_{k=1}^{n} \left\{ (1+u^2) -2u \cos \xi_k \right\}
= \prod_{k=1}^{n} \left(1 - e^{i \xi_k} u \right) \left(1 - e^{-i \xi_k} u \right) \\
&= \left\{ \prod_{k=1}^{n} \left(1 - e^{i \xi_k} u \right) \right\}^2 = \left( 1-u^n \right)^2.
\end{align*}
Thus we have
\begin{align}
\zeta_{{\bf U}_{C_n}} \left( u \right) = \frac{1}{\left( u^n - 1 \right)^2}.
\label{kirishima50}
\end{align}
Then we see that Eq. \eqref{kirishima50} is the same as Eq. \eqref{kirishima02}. Therefore it follows from  Eqs. \eqref{kirishima05} and \eqref{kirishima06} that $Z_{\zeta_{{\bf U}_{C_n}}} (w,s)$ and $\zeta_{\zeta_{{\bf U}_{C_n}}} (s)$ can be expressed as a Hurwitz zeta function of order $2$, $\zeta_2 (s, x, (\omega_1, \omega_2))$, and a gamma function of order $2$, $\Gamma_2 (x, (\omega_1, \omega_2))$, in the following way:
\begin{prop}
\begin{align*}
Z_{\zeta_{{\bf U}_{C_n}}} (w,s) &= \zeta_2 (w, s+2n, (n,n)), 
\\
\zeta_{\zeta_{{\bf U}_{C_n}}} (s) &= \Gamma_2 (s+2n, (n,n)). 
\end{align*}
\label{hokuseihou03}
\end{prop}
As for the proof, see \cite{Konno2023}.
\par
\
\par
(ii) Star graph case

Let $G=S_n$ be the star graph with $n$ vertices and $m=n-1$ edges. Remark that $S_n$ is isomorphic to the complete bipartite graph $K_{1,n-1}$. Then we obtain
\begin{align*}
{\rm Spec} \left( {\bf P}_{S_n} \right) 
&={\rm Spec} \left( {\bf P}_n \right) 
= \left\{ [1]^1, \ [0]^{n-2}, \ [-1]^1 \right\}, 
\\
{\rm Spec} \left( {\bf U}_{S_n} \right)
&={\rm Spec} \left( {\bf U}_{2(n-1)} \right)
= \left\{ [1]^2, \ \left[ i \right]^{n-2}, \ \left[ -i \right]^{n-2}, \ \left[ -1 \right]^2 \right\} \setminus \left\{ [1]^1, \ \left[ -1 \right]^1 \right\}.
\\
&=\left\{ [1]^1, \ \left[ i \right]^{n-2}, \ \left[ -i \right]^{n-2}, \ \left[ -1 \right]^1 \right\}. 
\end{align*}
From the definition of $\zeta_{{\bf U}_{S_n}} \left( u \right)$, we have 
\begin{align*}  
\zeta_{{\bf U}_{S_n}} \left( u \right) 
&= \left\{ \det \left( {\bf I}_{2n-2} - u {\bf U}_{S_n} \right) \right\}^{-1}
= \prod_{k=1}^{2n-2} \frac{1}{1 - \lambda_k \ u}
\\
&
= \frac{1}{1-u} \times  \frac{1}{1+u} \times \left(\frac{1}{1-iu} \right)^{n-2} \times \left(\frac{1}{1+iu} \right)^{n-2}  
\\
&
= \frac{1}{1-u^2} \times \left( \frac{1}{1+u^2} \right)^{n-2}
= \frac{1}{1-u^2} \times \left( \frac{1-u^2}{1-u^4} \right)^{n-2}
\\
&
= \frac{(1-u^2)^{n-3}}{(1-u^4)^{n-2}} = (-1) \times \frac{(u^2-1)^{n-3}}{(u^4-1)^{n-2}}.
\end{align*}
Then we get the following result.
\begin{align*}
\zeta_{{\bf U}_{S_n}} \left( x \right) =  (-1) \times \frac{(x^2-1)^{n-3}}{(x^4-1)^{n-2}}.
\end{align*}
 Moreover $\zeta_{{\bf U}_{S_n}}$ is the form that we can apply Theorem 4. Therefore we obtain $Z_{\zeta_{\mathbf{U}_{S_n}}}$ and $\zeta_{\zeta_{\mathbf{U}_{S_n}}}$ in the same way as Proposition 1.
 \begin{prop}
\begin{align*}
Z_{\zeta_{{\bf U}_{S_n}}} (w, s)
&= \sum_{I \subset \{1, \ldots, n-3 \}} (-1)^{|I|+1} \zeta_{n-2} \left( w, s + 2(n-1) + 2 |I|, \bec{n} \right),
\\
\zeta_{\zeta_{{\bf U}_{S_n}}} (s) 
&= \prod_{I \subset \{1, \ldots, n-3 \}} \Gamma_{n-2} \left( s +  2(n-1) + 2 |I|, \bec{n} \right)^{ (-1)^{|I|+1}}.
\end{align*}
\end{prop}
Here we denote $f$ by $\zeta_{{\bf U}_{S_n}}$ in Theorem 4. Noting that 
\begin{align*}
&\ell =0, \quad a=n-3, \quad m(1)= \cdots =m(n-3)=2, 
\\
&b=n-2, \quad n(1)= \cdots =n(n-2)=4, \quad \bec{n} = \overbrace{(4,\ldots,4)}^{n-2}, \quad |\bec{n}| = 4(n-2),
\end{align*}
it follows from Eq. \eqref{mkusatsu01b} that
\begin{align*}
Z_{\zeta_{{\bf U}_{S_n}}} (w, s)
= \sum_{I \subset \{1, \ldots, n-3 \}} (-1)^{|I|+1} \zeta_{n-2} \left( w, s + 2(n-1) + 2 |I|, \bec{n} \right).
\end{align*}
Similarly, by Eq. \eqref{mkusatsu02b}, we see 
\begin{align*}
\zeta_{\zeta_{{\bf U}_{S_n}}} (s) 
= \prod_{I \subset \{1, \ldots, n-3 \}} \Gamma_{n-2} \left( s +  2(n-1) + 2 |I|, \bec{n} \right)^{ (-1)^{|I|+1}}.
\end{align*}

\par
\
\par
(iii) Complete graph case

Let $G=K_n$ be the complete graph with $n$ vertices and $m=n(n-1)/2$ edges for $n \ge 4$. If $n=3$, then this graph belongs to the class of cycle graphs. If $n=2$, then this is one of the star graphs. Then we get
\begin{align*}
{\rm Spec} \left( {\bf P}_{K_n} \right) 
&= {\rm Spec} \left( {\bf P}_n \right) 
= \left\{ [1]^{1}, \  \left[ - \frac{1}{n-1} \right]^{n-1} \right\}, 
\\
{\rm Spec} \left( {\bf U}_{K_n} \right) 
&= {\rm Spec} \left( {\bf U}_{n(n-1)} \right)
= \Biggl\{ [1]^{(n(n-3)+4)/2}, \  [-1]^{n(n-3)/2}, \  \left[ \alpha \right]^{n-1}, \ \left[ \bar{\alpha} \right]^{n-1} \Biggr\},
\end{align*}
where 
\begin{align*}
\alpha = \frac{-1 + i \sqrt{n(n-2)}}{n-1}
\end{align*}
and $\bar{z}$ is the complex congugate of $z \in \mathbb{C}$. Put $L=n(n-3)/2$. The definition of $\zeta_{{\bf U}_{K_n}} \left( u \right)$ implies
\begin{align*}  
\zeta_{{\bf U}_{K_n}} \left( u \right) 
&= \left\{ \det \left( {\bf I}_{n(n-1)} - u {\bf U}_{K_n} \right) \right\}^{-1}
= \prod_{k=1}^{n(n-1)} \frac{1}{1 - \lambda_k \ u}
\\
&
= \frac{1}{(1-u)^{L+2}} \times  \frac{1}{(1+u)^L} \times \frac{1}{(1-\alpha u)^{n-1}} \times \frac{1}{(1- \bar{\alpha} u)^{n-1}}  
\\
&
= \frac{1}{(1-u)^{L+2}} \times  \frac{(1-u)^L}{(1-u^2)^L} \times \frac{1}{\left\{ 1- \left(\alpha + \bar{\alpha}\right) u + u^2 \right\}^{n-1}}
\\
&
= \frac{1}{(1-u)^{2}} \times  \frac{1}{(1-u^2)^L} \times \left( 1 + \frac{2}{n-1} u + u^2 \right)^{-(n-1)}
\\
&
= (-1)^L \times \frac{1}{(u-1)^{2}} \times \frac{1}{(u^2-1)^L} \times \left( u^2 + \frac{2}{n-1} u + 1 \right)^{-(n-1)}.
\end{align*}
In this case, $\zeta_{{\bf U}_{K_n}}$ does not satisfy the assumption of Theorem 4. Therefore we consider derivation of $Z_{\zeta_{\mathbf{U}_{K_n}}}$ and $\zeta_{\zeta_{\mathbf{U}_{K_n}}}$ from its definition. We deform the parts $\frac{1}{(u-1)^{2}}$, $\frac{1}{(u^2-1)^L}$, and $\left( u^2 + \frac{2}{n-1} u + 1 \right)^{-(n-1)}$ as follows.\\

First, we find \[\frac{1}{(u-1)^2} = u^{-2} \left( \sum_{k=0}^{\infty} u^{-k}\right)^2 = \left(\sum_{k=0}^{\infty} u^{-k-1}\right)^2\] and \[\frac{1}{(u^2-1)^L} = u^{-2L}\left(\sum_{r=0}^{\infty} u^{-2r}\right)^L = \left(\sum_{r=0}^{\infty} u^{-2r-2}\right)^L \]
by the binomial theorem.
Next, we consider the series transformation of $\left( u^2 + \frac{2}{n-1} u + 1 \right)^{-(n-1)}$. Let $\alpha$ and $\beta$ be the solutions of the equation of $1+\frac{2}{n-1}u + u^2=0$. Then $\alpha$ and $\beta$ satisfy the relation $\alpha + \beta = -\frac{2}{n-1}$ and $\alpha\beta=1$. Therefore we get 
\begin{align*}
    \left(1+\frac{2}{n-1}u + u^2\right)^{-(n-1)} 
    &=(u-\alpha)^{-(n-1)}(u-\beta)^{-(n-1)}\\
    &= u^{-2(n-1)}\left(1-\frac{\alpha}{u}\right)^{-(n-1)}\left(1-\frac{\beta}{u}\right)^{-(n-1)}\\
    &= u^{-2(n-1)}\left(\sum_{k=0}^{\infty} \alpha^k u^{-k}\right)^{n-1} \left(\sum_{r=0}^{\infty} \beta^r u^{-r}\right)^{n-1}\\
    &= u^{-2(n-1)}\left(\sum_{k=0}^{\infty} \sum_{r=0}^{\infty} \alpha^k \beta^r u^{-(k+r)}\right)^{n-1} \\
    &= u^{-2(n-1)}\left(\sum_{l=0}^{\infty} u^{-l} \sum_{\{(k,r) \in \mathbb{Z}_+^2 ; k+r=l \}}\alpha^{k}\beta^{r}\right)^{n-1}.
\end{align*}
Here $\mathbb{Z}_{+} = \{0,1,2,...\}.$
The part of $\sum_{\{(k,r) \in \mathbb{Z}_+^2 ; k+r=l \}}\alpha^{k}\beta^{r}$ is the symmetric expression, so this series is represented by the elementally symmetric expression.
\[\sum_{\{(k,r) \in \mathbb{Z}_+^2 ; k+r=l \}}\alpha^{k}\beta^{r}=\sum_{i=0}^{\lfloor \frac{l}{2} \rfloor}(-1)^i
\begin{pmatrix}
l-i \\
i \\
\end{pmatrix}
(\alpha \beta)^i \left(\alpha + \beta\right)^{l-2i}.\]
 For simplicity, we put 
 \[P_l = \sum_{i=0}^{\lfloor \frac{l}{2} \rfloor}(-1)^{l-i}
\begin{pmatrix}
l-i \\
i \\
\end{pmatrix}
 \left(\frac{2}{n-1}\right)^{l-2i}.\]
 Then we obtain 
 \begin{align*}
     \left(1+\frac{2}{n-1}u + u^2\right)^{-(n-1)} 
     &= u^{-2(n-1)}\left(\sum_{l=0}^{\infty} u^{-l}\sum_{i=0}^{\lfloor \frac{l}{2} \rfloor}(-1)^{l-i}
     \begin{pmatrix}
      l-i \\
      i \\
     \end{pmatrix}
     (\alpha \beta)^i (\alpha + \beta)^{l-2i}\right)^{n-1}\\
     &= u^{-2(n-1)}\left(\sum_{l=0}^{\infty} u^{-l}\sum_{i=0}^{\lfloor \frac{l}{2} \rfloor}(-1)^{l-i}
     \begin{pmatrix}
      l-i \\
      i \\
     \end{pmatrix}
    \left(\frac{2}{n-1}\right)^{l-2i}\right)^{n-1}\\
     &=  u^{-2(n-1)}\left(\sum_{l=0}^{\infty} P_l u^{-l}\right)^{n-1} \\
     &= \left(\sum_{l=0}^{\infty} P_l u^{-l-2}\right)^{n-1}.
 \end{align*}
The above argument leads 
 \begin{align*}
      \zeta_{U_{K_n}}(u) &= (-1)^{L}\left(\sum_{k=0}^{\infty} u^{-k-1}\right)^2 \left(\sum_{r=0}^{\infty} u^{-2r-2}\right)^L \left(\sum_{l=0}^{\infty} 
 P_l u^{-l-2}\right)^{n-1}\\
&=(-1)^L\sum_{k_1=0,k_2=0,...,k_{M}=0}^{\infty} \left(\prod_{N=L+3}^{M}{P_{k_N}}\right) u^{-(k_1+k_2+2)-\sum{(2k_m+2)}-\sum{(k_N+2)}}\\
&=(-1)^L\sum_{\bec{k} \geq 0} \left(\prod_{N=L+3}^{M}{P_{k_N}}\right) u^{-(k_1+k_2+2)-\sum{(2k_m+2)}-\sum{(k_N+2)}},
 \end{align*}
 where $M=\frac{n^2-n+2}{2}$ and $\bec{k}=(k_1,k_2,...,k_M)$. Therefore the absolute Hurwitz zeta function $Z_{\zeta_{U_{K_n}}(u)}$ is computed as follows.
 \begin{align}
    &Z_{\zeta_{U_{K_n}}}(w,s) \notag \\
    &= \frac{(-1)^L}{\Gamma(w)}\int_{1}^{\infty}\sum_{\bec{k} \geq 0} \left(\prod_{N=L+3}^{M}{P_{k_N}}\right) x^{-(k_1+k_2+2)-\sum{(2k_m+2)}-\sum{(k_N+2)}-s-1} (\log x)^{w-1}dx \notag \\
    &=\frac{(-1)^L}{\Gamma(w)}\int_{0}^{\infty}\sum_{\bec{k} \geq 0} \left(\prod_{N=L+3}^{M}{P_{k_N}}\right) e^{(-(k_1+k_2+2)-\sum{(2k_m+2)}-\sum{(k_N+2}))t-st} t^{w-1}dt \label{M.t}\\
    &= (-1)^L\sum_{\bec{k} \geq 0} \left(\prod_{N=L+3}^{M}{P_{k_N}}\right) \frac{1}{\Gamma(w)}\int_{0}^{\infty} e^{(-(k_1+k_2+2)-\sum{(2k_m+2)}-\sum{(k_N+2}))t-st} t^{w-1}dt \label{exchange}\\
    &= (-1)^L\sum_{\bec{k} \geq 0} \left(\prod_{N=L+3}^{M}{P_{k_N}}\right) \frac{1}{((k_1+k_2+2)+\sum{(2k_m+2)}+\sum{(k_N+2})+s)^w}. \notag
    \end{align}
Here, we use the Mellin transform in Eq. (\ref{M.t}). In addition, we get the fact $\{P_l\}$ is bounded, so the order of sum and integration in Eq. (\ref{exchange}) can exchange. Then we derive the {\em absolute zeta function} $\zeta_{\zeta_{U_{K_n}}}$ in the following fashion.
\begin{align}
    &\zeta_{\zeta_{U_{K_n}}}(s) \notag\\
    &= \exp \left[\frac{\partial}{\partial w}\sum_{\bec{k} \geq 0} \left(\prod_{N=L+3}^{M}{P_{k_N}}\right) \frac{(-1)^L}{((k_1+k_2+2)+\sum{(2k_m+2)}+\sum{(k_N+2})+s)^w}|_{w=0}\right] \label{a.z1}\\
    &= \exp \left[ \sum_{\bec{k} \geq 0} (-1)^L\left(\prod_{N=L+3}^{M}{P_{k_N}}\right) \log \left(\frac{1}{(k_1+k_2+2)+\sum{(2k_m+2)}+\sum{(k_N+2})+s}\right)\right]\label{a.z2}\\
    &= \exp\left[\sum_{\bec{k} \geq 0} \log \left(\frac{1}{(k_1+k_2+2)+\sum{(2k_m+2)}+\sum{(k_N+2})+s}\right)^{(-1)^L\prod_{N=L+3}^{M}{P_{k_N}}}\right]\label{a.z3}\\
    &= \prod_{k_1,...,k_{M}} ((k_1+k_2+2)+\sum{(2k_m+2)}+\sum{(k_N+2)})+s)^{(-1)^L\prod_{N=L+3}^{M}{P_{k_N}}} .\label{a.z4}
\end{align}
The series $\sum_{\bec{k} \geq 0} (\prod_{k_N}{P_{k_N}}) ((k_1+k_2+2)+\sum{(2k_m+2)}+\sum{(k_N+2})+s)^{-w}$ diverges to infinity around the zero point.
However $Z_f(w,s)$ has an analytic continuation that is differentiable in the zero point, so the partial differential in Eqs. (\ref{a.z1}), (\ref{a.z2}), (\ref{a.z3}), and (\ref{a.z4}) means the the partial differential of analytic continuation of $Z_f(w,s)$. Therefore we get $Z_{\zeta_{U_{K_n}}(u)}$ and $\zeta_{\zeta_{U_{K_n}}(u)}$ as the analogue of the multiple Hurwitz zeta function and the multiple gamma function.
 \begin{prop}
\begin{align*}
Z_{\zeta_{{\bf U}_{K_n}}} (w, s)
&= (-1)^L \sum_{\mathbf{k} \geq 0}  \frac{\prod_{N=L+3}^{M}{P_{k_N}}}{((k_1+k_2+2)+\sum{(2k_m+2)}+\sum{(k_N+2})+s)^w} , 
\\
\zeta_{\zeta_{{\bf U}_{K_n}}} (s) 
&=  \prod_{k_1,...,k_{M}} ((k_1+k_2+2)+\sum{(2k_m+2)}+\sum{(k_N+2)}+s)^{(-1)^L\prod_{N=L+3}^{M}{P_{k_N}}}.
\end{align*}
\end{prop}

(iv) Complete bipartite graph case.

Let $K_{n_1, n_2}$ be the complete bipartite graph. Then $K_{n_1,n_2}$ has $n = n_1 + n_2$ vertices and $m= n_1 n_2$ edges. The class of complete bipartite graphs includes star graphs which  hold $n_1=1, \ n_2=n-1$  as a special case. We obtain the following eigenvalues. See Higuchi et al. \cite{HiguchiEtAl2017}, for example.
\begin{align*}
&{\rm Spec} \left( {\bf P}_{K_{n_1, n_2}} \right) 
={\rm Spec} \left( {\bf P}_{n_1 + n_2} \right) 
= \left\{ [1]^1, \ [0]^{n_1 + n_2 -2}, \ [-1]^1 \right\}, 
\\
&{\rm Spec} \left( {\bf U}_{K_{n_1, n_2}} \right)
={\rm Spec} \left( {\bf U}_{2 n_1 n_2} \right)
\\
& \quad =\left\{ [1]^{n_1 n_2 -(n_1 + n_2) + 2}, \ \left[ i \right]^{n_1 + n_2 - 2}, \ \left[ -i \right]^{n_1 + n_2 - 2}, \ \left[ -1 \right]^{n_1 n_2 -(n_1 + n_2) + 2} \right\}. 
\end{align*}
Then we derive $\zeta_{{\bf U}_{K_{n_1, n_2}}} \left( u \right)$ by ${\rm Spec} \left( {\bf U}_{K_{n_1, n_2}} \right)$ in the following way.
\begin{align*}  
&\zeta_{{\bf U}_{K_{n_1, n_2}}} \left( u \right) 
\\
&= \left\{ \det \left( {\bf I}_{2 n_1 n_2} - u {\bf U}_{K_{n_1, n_2}} \right) \right\}^{-1}
= \prod_{k=1}^{2 n_1 n_2} \frac{1}{1 - \lambda_k \ u}
\\
&
= \left( \frac{1}{1-u} \right)^{n_1 n_2 -(n_1 + n_2) + 2} \times \left( \frac{1}{1+u} \right)^{n_1 n_2 -(n_1 + n_2) + 2} 
\\
& \qquad \times \left(\frac{1}{1-iu} \right)^{n_1 + n_2 - 2} \times \left(\frac{1}{1+iu} \right)^{n_1 + n_2 - 2}  
\\
&
= \left( \frac{1}{1-u^2} \right)^{n_1 n_2 -(n_1 + n_2) + 2} \times \left( \frac{1}{1+u^2} \right)^{n_1 + n_2 - 2}
\\
&= \left( \frac{1}{1-u^2} \right)^{n_1 n_2 -(n_1 + n_2) + 2} \times \left( \frac{1-u^2}{1-u^4} \right)^{n_1 + n_2 - 2}
\\
&
= \frac{(1-u^2)^{ 2 ( n_1 + n_2 - 2 ) - n_1 n_2 }}{(1-u^4)^{n_1 + n_2 - 2}} \\
&= (-1)^{n_1 + n_2 - n_1 n_2} \times \frac{(u^2-1)^{ 2 (n_1 + n_2 - 2) - n_1 n_2 }}{(u^4-1)^{n_1 + n_2 - 2}}.
\end{align*}
Thus we get 
\begin{align}
\zeta_{{\bf U}_{K_{n_1, n_2}}} \left( u \right) = (-1)^{n_1 + n_2 - n_1 n_2} \times \frac{(u^2-1)^{ - (n_1 -2)(n_2 - 2)}}{(u^4-1)^{n_1 + n_2 - 2}}.
\label{hakuouhounibu}
\end{align}
According to Eq. \eqref{hakuouhounibu}, we have 
\begin{align*}
\zeta_{{\bf U}_{K_{n_1, n_2}}} \left( \frac{1}{u} \right) = (-1)^{n_1 + n_2 - n_1 n_2} \cdot u^{2 n_1 n_2} \cdot \zeta_{{\bf U}_{K_{n_1, n_2}}} (u).
\end{align*}
Then we find $\zeta_{{\bf U}_{K_{n_1, n_2}}} \left( u \right)$ is the absolute automorphic form of weight $-2 n_1 n_2$. 
We assume that $n_1 + n_2 \ge 2$ and $1 \le n_1 \le n_2$, where $n_1, \ n_2 \in \ZM_{>}$. The case of $n_1 = 1, \ n_2 = n-1 \ (n \ge 2)$ is one of the example that we mentioned in (ii). Moreover, we classify this case into three cases.

\par
\
\par
(iv-a) The case of $(n_1, n_2) = (1, n_2)$, where $n_2 = 3, 4, 5, \ldots$. Then we get 
$- (n_1 -2)(n_2 - 2) \ge 1$, because $n_2 - 2 \ge 1$ holds and $n_1 = 1$. Thus, we obtain $n_2 \ge 3$. The zeta function $\zeta_{{\bf U}_{K_{1, n_2}}}$ is represented as follows.
\begin{align}
\zeta_{{\bf U}_{K_{1, n_2}}} \left( u \right) = (-1) \times \frac{(u^2-1)^{n_2-2}}{(u^4-1)^{n_2 - 1}}.
\label{hakuouhounibu4a}
\end{align}
Then we obtain the absolute Hurwitz zeta function and {\em absolute zeta function} by Theorem \ref{zettaisugaku01chap2}. Noting that the right side of Eq. \eqref{hakuouhounibu4a} has a factor of $(-1)$, we obtain
\begin{prop}
    \begin{align*}
&Z_{\zeta_{{\bf U}_{K_{1, n_2}}}} (w, s)
= \sum_{I \subset \{1, \ldots, n_2-2 \}} (-1)^{|I|+1} \zeta_{n_2-1} \left( w, s + 2 n_2 + 2 |I|, \bec{n} \right),
\\
&\zeta_{\zeta_{{\bf U}_{K_{1, n_2}}}} (s) 
= \prod_{I \subset \{1, \ldots, n_2-2 \}} \Gamma_{n_2-1} \left( s + 2 n_2 + 2 |I|, \bec{n} \right)^{ (-1)^{|I|+1}},
\\
&\zeta_{\zeta_{{\bf U}_{K_{1, n_2}}}} (-2 n_2 -s)^{-1} = \varepsilon_{\zeta_{{\bf U}_{K_{1, n_2}}}} (s) \ \zeta_{\zeta_{{\bf U}_{K_{1, n_2}}}} (s), 
\end{align*}
when $\ell =0, \ a= n_2-2, \ m(1)= \cdots = m(a)=2, \ b=n_2-1, \ n(1)= \cdots =n(n_2-1)=4, \ {\rm deg} (f) = D =-2 n_2, \ C=-1$, and  
\begin{align*}
&\bec{n} 
= (\overbrace{4, \ldots, 4}^{n_2-1}), \qquad
\\ 
&\varepsilon_{\zeta_{{\bf U}_{K_{1, n_2}}}} (s) = 
\prod_{I \subset \{1, \ldots, n_2-2 \}} S_{n_2-1} \left(s + 2 n_2 + 2 |I|, \bec{n} \right)^{ (-1)^{|I|+1}}.
\end{align*}
\end{prop}
Here $S_r(\cdot,\cdot)$ is the multiple sine function. Concerning the relation of the multiple gamma function and the multiple sine function, see \cite{Kurokawa3, Kurokawa, KT3}, for instance.
\par
\
\par
(iv-b) The case of $(n_1, n_2) = (1,2), \ (2, n_2)$, where $n_2 = 2, 3, 4, \ldots$. The equation $- (n_1 -2)(n_2 - 2) = 0$ can be easily verified because $n_1=2$ or $n_2=2$. Thus, the zeta function of this case $\zeta_{{\bf U}_{K_{n_1, n_2}}} $ is represented as follows, because $n_1 n_2$ is even.
\begin{align}
\zeta_{{\bf U}_{K_{n_1, n_2}}} \left( u \right) = \frac{(-1)^{n_1 + n_2}}{(u^4-1)^{n_1 + n_2 - 2}}.
\label{hakuouhounibu4b}
\end{align}
Then we apply Theorem {\rm \ref{zettaisugaku01chap2}} in this case. Here we note that the right side of Eq. \eqref{hakuouhounibu4b} has a factor $(-1)^{n_1 + n_2}$.
\begin{prop}
    \begin{align*}
&Z_{\zeta_{{\bf U}_{K_{n_1, n_2}}}} (w, s)
= (-1)^{n_1 + n_2}\zeta_{n_1+n_2-2} \left( w, s + 4 (n_1 + n_2 - 2), \bec{n} \right),
\\
&\zeta_{\zeta_{{\bf U}_{K_{n_1, n_2}}}} (s) = \Gamma_{n_1+n_2-2} \left( s + 4 (n_1 + n_2 - 2), \bec{n} \right)^{ (-1)^{|I|+n_1 + n_2}},
\\
&\zeta_{\zeta_{{\bf U}_{K_{n_1, n_2}}}} (-4 (n_1 + n_2 - 2)-s)^{{(-1)}^{n_1 + n_2}} = \varepsilon_{\zeta_{{\bf U}_{K_{n_1, n_2}}}} (s) \ \zeta_{\zeta_{{\bf U}_{K_{n_1, n_2}}}} (s),
\end{align*}
where $\ell =0, \ a=0, \ b=n_1+n_2-2, \ n(1)= \cdots =n(n_1+n_2-2)=4, \ {\rm deg} (f) = D = -4(n_1 + n_2 - 2) , \  C=(-1)^{n_1 + n_2}$, and 
\begin{align*}
&\bec{n} 
= (\overbrace{4, \ldots, 4}^{n_1+n_2-2}), \qquad
\\ 
&\varepsilon_{\zeta_{{\bf U}_{K_{n_1, n_2}}}} (s) = 
S_{n_1+n_2-2} \left(s + 4 (n_1 + n_2 - 2), \bec{n} \right)^{ (-1)^{n_1 + n_2 }}.
\end{align*}
\end{prop}

\par
\
\par
(iv-c) The case of $(n_1, n_2) = (1,1), \ (3, n_2)$, where $n_2 =3, 4, 5, \ldots$. In this case, we get $- (n_1 -2)(n_2 - 2) \le -1$. In fact, $- (1 - 2)(1 - 2) = -1 \le -1$ holds if $(n_1, n_2) = (1,1)$, and $(n_1, n_2) = (3, n_2)$, then $- (n_2 - 2) \le - 1$. Therefore, we obtain $n_2 \ge 3$. Moreover we find $(-1)^{n_1 + n_2 - n_1 n_2} = -1$ with the fact that $n_1 \in \{1,3\}$ and $n_1 + n_2 - n_1 n_2 \equiv 1 \ ({\rm mod} \ 2)$. We get 
\begin{align*}
\zeta_{{\bf U}_{K_{n_1, n_2}}} \left( u \right) = - \frac{1}{(u^2-1)^{(n_1 -2)(n_2 - 2)}(u^4-1)^{n_1 + n_2 - 2}}.
\end{align*}
Then we derive the absolute Hurwitz zeta function and the absolute zeta function similarly to the above case. That is to say, we apply Theorem \ref{zettaisugaku01chap2} to $\zeta_{{\bf U}_{K_{n_1, n_2}}}$.
\begin{prop}
    \begin{align*}
&Z_{\zeta_{{\bf U}_{K_{n_1, n_2}}}} (w, s)
= - \ \zeta_{(n_1 -1)(n_2 -1)+1} \left( w, s + 2 n_1 n_2, \bec{n} \right),
\\
&\zeta_{\zeta_{{\bf U}_{K_{n_1, n_2}}}} (s) = \Gamma_{(n_1 -1)(n_2 -1)+1} \left( s + 2 n_1 n_2, \bec{n} \right)^{-1},
\\
&\zeta_{\zeta_{{\bf U}_{K_{n_1, n_2}}}} (- 2 n_1 n_2-s)^{-1} = \varepsilon_{\zeta_{{\bf U}_{K_{n_1, n_2}}}} (s) \ \zeta_{\zeta_{{\bf U}_{K_{n_1, n_2}}}} (s), 
\end{align*}
where $\ell =0, \ a=0, \ b=(n_1 -1)(n_2 -1)+1, \ b^{\ast} =(n_1 -2)(n_2 - 2), \ n(1)= \cdots =n(b^{\ast})=2, \  n(b^{\ast}+1)= \cdots =n(b)=4, \
{\rm deg} (f) = D = 2 n_1 n_2 , \  C=-1$, and 
\begin{align*}
&\bec{n} 
= (\overbrace{2, \ldots, 2}^{b^{\ast}},\overbrace{4, \ldots, 4}^{b - b^{\ast}}), \qquad
\\ 
&\varepsilon_{\zeta_{{\bf U}_{K_{n_1, n_2}}}} (s) = 
S_{(n_1 -1)(n_2 -1)+1} \left(s + 2 n_1 n_2, \bec{n} \right)^{-1}.
\end{align*}
\end{prop}

\section{Conclusion \label{sec06}}
In this paper, first we dealt with a zeta function $\zeta_{{\bf U}_{G}} (u)$ determined by ${\bf U}_{G}$. Here ${\bf U}_{G}$ is a time evolution matrix of the Grover walk on $G$ and $G$ is a simple connected graph with $n$ vertices and $m$ edges. Then we proved that $\zeta_{{\bf U}_{G}} (u)$ is an absolute automorphic form of weight $- 2m$ (Theorem \ref{kakumei09}). After that we considered the absolute zeta function $\zeta_{\zeta_{{\bf U}_{G}}} (s)$ for our zeta function $\zeta_{{\bf U}_{G}} (u)$. As examples, we treated some of simple connected graphs, especially cycle graph, star graph, complete graph, and complete bipartite graph. We introduced two ways of calculation of the absolute zeta function. The functions for cycle graph, star graph, and complete bipartite graph were computed by Theorem {\rm \ref{zettaisugaku01chap2}}. This is related to the period of the Grover walk on the graph. If a period of the Grover walk is finite, then the absolute zeta function derived from Theorem {\rm \ref{zettaisugaku01chap2}}. On the other hand, the function constructed by the Grover walk with infinite period, like complete graph, was computed by the Konno-Sato theorem (Theorem \ref{K-S}) and its definition. One of the interesting future problems might be to get $\zeta_{\zeta_{{\bf U}_{G}}} (s)$ for other graphs except for the cycle graph, star graph, and complete bipartite graph. This future work will lead to get the zero point of the absolute zeta function of our zeta function. Actually, the zero point of the absolute zeta function of the Grover walk can be obtained. We are interested in the property of the Grover walk derived from the distribution of the zero point.

\section*{Data Availability}
Not applicable.

\section*{Conflict of Interest}
Not applicable.

\par
\
\par
\noindent


\end{document}